\documentstyle[multicol,aps,epsfig]{revtex}

\newcommand{\bc}{\begin{center}}
\newcommand{\ec}{\end{center}}
\newcommand{\bg}{\begin{equation}}
\newcommand{\ed}{\end{equation}}

\newcommand{\kt}{k_{\rm B}T}
\newcommand{\lb}{l_{\rm B}}

\begin{document}

\title{Polyelectrolyte gels in poor solvent: Equilibrium and non equilibrium
  elasticity} \author{T.A. Vilgis$^{(1,2)}$, A. Johner$^{(1)}$, J. F.
  Joanny$^{(1)}$} 

\address{${^{(1)}}$Laboratoire Europ\'een Associ\'e, Institut Charles Sadron,
  \\6, rue Boussingault, F- 67083 Strasbourg, Cedex, France}
\address{$^{(2)}$Max-Planck-Institut f\"ur Polymerforschung, \\Postfach 3148,
  D-55021 Mainz, Germany}

\date{\today} 

\maketitle

\begin{abstract}
  We study theoretically using scaling arguments the behavior of
  polyelectrolyte gels in poor solvents. Following the classical picture of
  Katchalsky, our approach is based on single chain elasticity but it accounts
  for the recently proposed pearl-necklace structure of polyelectrolytes in
  poor solvents.  The elasticity both of gels at swelling equilibrium and of
  partially swollen, non equilibrium, gels is studied when parameters such as
  the ionic strength or the fraction of charged monomers are varied. Our
  theory could be useful to interpret recent experiments performed in
  Strasbourg that show that if identical gel samples are swollen to the same
  extent at different pH the sample with the highest charge has the lowest
  shear modulus.
\end{abstract}

\section{Introduction}

Polyelectrolyte gels can absorb a considerable amount of water and swell up to
1000 times.  This remarkable property can be monitored by the salt
concentration and makes polyelectrolyte gels good candidates for
superadsorbent materials.  From a fundamental point of view, polyelectrolyte
gels are less understood than neutral gels. In some respects, a neutral gel at
swelling equilibrium is similar to a semi-dilute polymer solution at the
overlap concentration, the mesh size in the gel grows up to the point where
the meshes marginally overlap, this is the so-called $c^\star$ description.
Neutral gels thus benefit to a large extent from the powerful scaling
description developed for single neutral chains and solutions. The situation
is, however, complicated by trapped entanglements that can, depending on the
preparation process, severely restrict swelling and by heterogeneities
(fluctuations in the crosslink density) that can be revealed by swelling
\cite{bastide,panyukov:98}.

There is no such strong analogy between polyelectrolyte gels and solutions,
mainly because of the long ranged electrostatic interaction and the presence
of counterions.  Counterions are free in the solution whilst they are trapped
in the equilibrium gel (immersed in a large excess of salt-free water) to
ensure macroscopic electroneutrality.  The osmotic pressure of the trapped
counterions, with no counterpart in solution, turns out to be the main reason
for swelling.  Though it originates from the electrostatic interaction between
polymer and counterions this pressure is not electrostatic in nature, it does
not depend on the strength of the electrostatic interaction, say on the
Bjerrum length. The forces responsible for the swelling may be identified as
the electrostatic forces acting on the gel boundary where the counterion
concentration vanishes.  As often in these matters, the thermodynamic point of
view, where the contribution from the narrow non neutral boundary is
negligible, is more convenient.

Neutral gels also appear simple in the sense that the chain interactions do
not play a major role in the elastic properties. The elastic modulus for
unentangled networks for example can be simply determined from the density of
crosslinks.  Even very refined theories \cite{edwards:88,panyukov:90} show
clearly that the modulus for untentangled networks is given by $G=c\kt/N$,
where $c$ is the monomer density and $N$ is the number of monomers of the
chain strand between two crosslinks.  This simple result is quite independent
of solvent quality; the excluded volume interactions influence mainly the bulk
properties such as the compression modulus, rather than the shear elastic
effects; the shear experiments probe, to first order, the strand elasticity
and the shear modulus is proportional to the elastic energy density stored in
the gel. The elasticity of polyelectrolyte gels is less simple because the gel
structure is not simply that of the polyelectrolyte solution. Due to the
stretching by the counterion pressure the shear modulus is actually higher
than what would be simply given by the crosslink density; it is further
expected to explicitly depend on the charge density, the quality of the
solvent and the salt concentration.

In the present paper we are concerned with the behavior of polyelectrolyte
gels in poor solvent.  Khokhlov and coworkers have studied in details the
behavior of such "responsive" gels on the basis of the Flory-Rehner theory,
i.e., the additivity of the different parts of the free energy.  The starting
point of these models is the Gaussian elastic free energy of the neutral
networks to which all the other contributions such as, the entropy of the
counterions, the electrostatic contributions or the solvent-monomer
interactions are added \cite{khokrev}.

The aim of the present paper is different. We study the elastic modulus of the
polyelectrolyte gel in a poor solvent using the recent work on the
conformation of polyelectrolyte chains.  Polyelectrolytes in poor solvents
are predicted to form pearl necklace structures
\cite{dobrynin:96,dobrynin:99}.  These structures result from a Rayleigh
instability of the collapsed globules if their radius exceeds a certain size.
In a previous paper we have studied the stretching of these pearl necklace
structures \cite{vilgis:99.5}. We calculated stress-strain relationships in
various situations. For chains with many pearls, we found a continuous
stress-strain relation.  If the necklace has only a few pearls, it stretches
discontinuously and the pearls are dissolved one by one. 

\begin{figure}
\begin{center}
\begin{minipage}{12cm}
\centerline{\epsfig{file=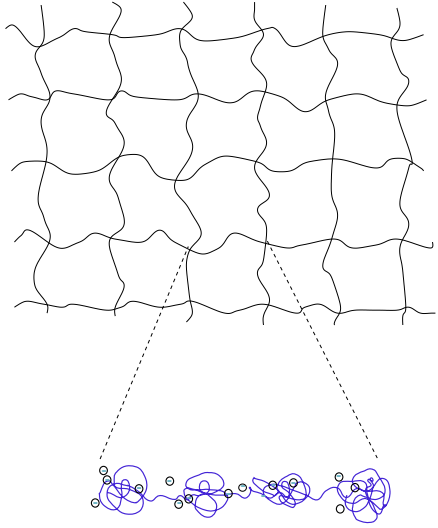,height=7cm}}
\vspace{10pt}
\caption{\label{fig:netw} The model of gel under investigation. The network
  consists of strands which can be described by polyelectrolyte chains in poor
  solvent. The strands themselves posses a pearl necklace structure}
\end{minipage}
\end{center}
\end{figure}

The model used in the paper below consists of weakly charged crosslinked
chains, with $N$ monomers of size $b$, a fraction $f$ of which are charged.
The interaction strength of the Coulomb potential $V({\bf r}) = \kt \lb /r$ is
expressed by the Bjerrum length $\lb = e^2/(4\pi \epsilon k_{\rm B}T)$;
$\epsilon$ is the dielectric constant of the solvent.

The paper is organized as follows: We first summarize the known results on
polyelectrolyte gels in $\theta$-solvent and on necklaces
\cite{dobrynin:96,vilgis:99.5}.  We then discuss equilibrium gels and
nonequilibrium gels in section III and finally present our conclusions.

\section{Polyelectrolyte gels in $\theta$-solvent and necklace
conformations }

In view of the following considerations it is useful to review first the main
points of the present state of the art, via simplest scaling arguments. These
include the theory of polyelectrolyte gels in $\theta$-solvent\cite{barrat:92},
the illustration of the role of the counterions, and the necklace description
of polyelectrolytes in poor solvents\cite{dobrynin:99,vilgis:99.5}. For the
latter we are going to discuss single chain elasticity and semi-dilute
solutions.

\subsection{Polyelectrolyte gels in $\theta$-solvent}

The size of an isolated polyelectrolyte chain in $\theta$-solvent can be
estimated from the balance between entropy and electrostatic energy.  A weakly
charged chain in a $\theta$-solvent stretches provided that the fraction of
charged monomers $f$ is large enough, $f>N^{-3/4}$. The conformation remains
locally isotropic at scales smaller than the electrostatic blob size
\begin{equation}
R_{\rm c} = b(l_{\rm B}f^2/a)^{-1/3}
\label{electblob}
\end{equation}
and the chain is extended at larger length scales.  This leads to the usual
blob model \cite{degennes:76} with an overall radius : $R = b N
\left(\frac{l_{\rm B}f^2}{b} \right)^{1/3}$ In solution, the chains overlap at
concentrations $c$ larger than the overlap concentration $c^\star = N/R^3$ .
In a semi-dilute solution ($c>c^\star$) the chains make a random walk of
step length $\xi_{\rm c} = \left(l_{\rm B}f^2/a\right)^{-1/6}(ca^3)^{-1/2}b$ with an
overall size
\begin{equation}
R_{1/2} = N^{1/2}\left(l_{\rm B} f^2/b\right)^{1/12}(cb^3)^{-1/4} b
\label{theta1/2}
\end{equation}
that becomes equal to the gaussian radius at high concentration when the
electrostatic blobs begin to overlap at concentrations $c\gg c^{**}\sim
f^{2/3}$.

The swelling equilibrium of a charged gel results from a balance between the
counterion contribution to the osmotic pressure $\Pi_{\rm counter} = k_{\rm
  B}Tfc$ and the elastic contribution $\Pi_{\rm elastic} = - k_{\rm B}T
cR^2/N^2b^2$, the equilibrium meshsize and concentration are:
\begin{equation}
R = Nf^{1/2}b \qquad c = N^{-2}f^{-3/2}b^{-3}
\end{equation}
where the so-called $c^{*}$ theorem $c = N/R^3$ which assumes disentangled
meshes is used.  The meshsize $R$ is larger than the size of an isolated chain
comprising $N$ monomers and does not depend on the Bjerrum length.  The
contribution of the direct electrostatic interaction to the energy density
$F_{\rm e} = k_{\rm B}Tl_{\rm B}f^2N^2/R^4$ can be checked to be negligible.

The addition of salt in water at low concentrations $n_{\rm s}<cf$ does not
affect the swelling. At higher salt concentrations, however, the gel is
expected to deswell, as the small ion contribution to the osmotic pressure in
the gel approaches the osmotic pressure in the surrounding bulk solution. The
ion partitioning is fixed by a Donnan equilibrium , that fixes the inner salt
concentration $n_{si}$ and to the reduced small ion pressure difference
$\Delta\Pi_{\rm ion} = c^2f^2/(4 n_{\rm s})$.  This pressure difference is
compensated by the elastic contribution of the chains in an equilibrium gel
and the concentration is
\begin{equation}
c = N^{-4/5}f^{-6/5}(n_{\rm s}b^3)^{3/5}b^{-3}
\label{thetasalt}
\end{equation}
It is easily checked that the direct electrostatic energy density remains
negligible as long as the mesh is weakly screened $\kappa R<1$. In the strong
screening limit $\kappa R>1$ the interactions become short-range. If we impose
a scaling form $F_{\rm e} = k_{\rm B}T(l_{\rm B}f^2N^2/R^4) {\rm f}(\kappa R)$
and require that the free energy density is quadratic in the polymer charge
density we obtain ${\rm f}(x) \propto x^{-2}$ for the scaling function and the
electrostatic free energy $F_{\rm e}$ is of the same order as $\Delta\Pi_{\rm ion}$.
Equation(\ref{thetasalt}) is thus expected to hold at the scaling level.

\subsection{Necklace structure and elasticity}

In a poor solvent a long neutral polymer chain collapses into a dense globule.
The poor
solvent conditions are
characterized by a negative excluded volume $v = -
\tau b^3$, where $\tau =
(\theta -T)/\theta$ is the relative distance to the Flory compensation
temperature $\theta$. In the absence of electrostatic interactions, the chain
collapses into a dense globule that may be
viewed as a small region of dense polymer phase at co-existence;
the finite size of the globule is associated with an extra energy
penalty due to the polymer-water
surface tension. The balance of the osmotic pressure between  the dense and
dilute phases (the latter at almost
vanishing concentration)  yields  the concentration inside the globule
 $c =
\tau / b^3$. Alternatively, the dense phase may be described by a close
packing of thermal blobs of
size
$\xi_t = b/\tau$, containing
$g=1/\tau^2$ monomers. A globule of $N$ monomers has then a radius $R =
b \left( N/\tau
\right)^{1/3}$. The corresponding surface tension is $\gamma = k_{\rm
B}T/\xi_t^2 \propto
\tau^2/b^2$.

In a poor solvent, a long polyelectrolyte chain adopts an elongated shape
determined by a balance between surface tension and electrostatic self-energy.
The first model was proposed by Kokhlov who optimized the free energy of a
cylindrical globule to obtain the transverse radius given by the electrostatic
blob radius $R_{\rm c}$ eq.(\ref{electblob}), and the length $L$ imposed by
the globule volume $V = N b^3/\tau \sim LR_{c}^{2}$. At least at the scaling
level, this description holds as long as there is no additional length scale.
Recently it was suggested that the charged globule is subject to a Rayleigh
instability and splits into connected droplets, the pearls, with a size of
order $R_{\rm c}$. The polymer strands separating the pearls are stretched due
to the electrostatic repulsion. However equilibrium between pearls and strands
imposes the strand tension.  In a first approximation, the strand tension is
the equilibrium tension $k_{\rm B}T\tau/b$ when a polymer strand is pulled out
of the dense equilibrium phase; the strands can then be viewed as linear
arrays of Gaussian blobs of size $\xi_{t}=b/\tau$. Noting that, up to a
logarithmic factor, the repulsion between two half-necklaces is the same as
the interaction between two adjacent pearls, we find the strand length $l$ and
the total chain length $L$.
\begin{equation}
l = b(b\tau/l_{\rm B}f^2)^{1/2}\qquad L = b N(l_{\rm B}f^2/b)^{1/2}\tau^{-1/2}
\; ,
\label{eqneck}
\end{equation}
where we use the pearl mass $ m = \tau b/(l_{\rm B} f^2)$ associated with
the radius $R_{\rm c}$.

At low charge fraction or/and very poor solvent, the pearl surface potential
calculated in the previous picture is larger than the thermal energy. The
counterions are regulated by the pearls, this happens for $\tau>\alpha^3
(bf/l_{\rm B})^{1/3}$ with $-\alpha \kt$ the counterion chemical potential.
The condensation of counterions inside the globules actually suppresses the
Rayleigh instability and a large globule is stable against the necklace.

Added salt at concentration $n_{\rm s}$, screens the electrostatic
interactions, the screening length $\kappa^{-1}$ is related to the salt
concentration through $\kappa^2 = 8\pi\l_{\rm B}n_{\rm s}$.  The necklace
structure itself is only marginally affected as long as the interaction
between adjacent pearls is not screened $\kappa l<1$.  For higher salt
concentrations however the distance between nearest pearls along the chain is
almost fixed at the screening length.

In a salt-free solution, necklaces overlap above the concentration $c^\star =
N/R^3$ and a single chain can be viewed as a random walk with a step length
$\xi_{\rm c} = (l_{\rm B}f^2/b)^{-1/4}\tau^{1/4}(c b^3)^{-1/2}b$, independent
of the overall chain size, and a radius\cite{dobrynin:99}:
\begin{equation}
R = N^{1/2}(l_{\rm B} f^2/b)^{1/8}(c b^3)^{-1/4}\tau^{-1/8} b
\label{1/2dilneck}
\end{equation}
This description holds at moderate concentration as long as  there are
several pearls per correlation
length $\xi_{\rm c}$, at higher concentration there remains one pearl per
correlation length; we not consider this regime any longer.

When the single necklace is submitted to an external pulling force $\varphi$,
pearls are progressively dissolved and converted into strand when the force is
increased. When there are only a few pearls, each pearl dissolution translates
into a length jump under imposed force or a force drop under imposed length.
For a higher number of pearls however the pearl number fluctuates by more than
one unit, single pearl features are washed out and the force curve is
continuous.  This is the regime of interest for the gel system where
elastically active paths can be quite long. The chain length can be related to
the pulling force by a force balance on the half-necklace similar to the one
leading to eq.(\ref{eqneck}) including the external force $\varphi$:
\begin{equation}
L(\varphi) = L(0)\left(1-{\varphi b\over \kt\tau}\right)^{-1/2}
\end{equation}
(where all finite size corrections are neglected).  The elastic energy stored
in the chain is dominated by the stretched strands (provided $L>L(0)$) and is
about $k_{\rm B}T$ per tensile-blob of radius $\xi_t$. This simple argument
fails when most of the pearls are dissolved, i.e. for $L>Nb\tau$.The maximum
length (when all pearls are used up) which represents an upper bound for the
validity of the continuous model.

\section{Polyelectrolyte gels in poor solvent}
\subsection{Equilibrium gels}

We want here to include the details the pearl-necklace structure for the
elastic chains of the gel; this cannot be done using complete calculations as
for the $\theta$-gel. However, useful predictions can be made at the scaling
level. We propose two different models : the $c^\star$-gel where the meshes
are disentangled at swelling equilibrium and for which the preparation state
is irrelevant, and the affine deformation model where the equilibrium swelling
depends on the preparation state. The affine model considers that the
deformation of the gel with respect to the preparation state is affine at all
length scales; this imposes the constraint that the polymer content of a mesh
volume remains constant upon swelling $c L_m^3/N = c_{\rm p}L_{\rm
  p}^3/N=\alpha$ where index $p$ refers to the preparation state. This
constraint can be understood as imposing that initially entangled meshes do
not disentangle. The special (limiting) case $\alpha = 1$ corresponds to the
$c^\star$-gel, obtained here by reacting all temporary contact points between
different chains of a semi-dilute solution during the crosslinking process.
Higher values of $\alpha$ are obtained if the gel is prepared in a semidilute
or a dense solution when only a small fraction of the temporary contact points
are crosslinked.

\subsubsection{$c^\star$-model}

The $c^\star$-model is based on the Katchalsky picture \cite{katchalsky:51}.
The equilibrium gel is immersed in a solvent reservoir that imposes its
osmotic pressure. In the absence of salt (actually if the salt concentration
$n_{s}$ is such that $n_{\rm s}<cf$), the osmotic pressure in the gel vanishes
(it is much lower than the ideal gas counterion pressure $\kt cf$). The
contribution arising from the counterions and from the strand elasticity thus
(almost) cancel.
\begin{equation}
\Pi = \Pi_{\rm elastic} + \Pi_{\rm count} = 0
\end{equation}
The counterion contribution to the osmotic pressure, similar to that for the
$\theta$-gel is dominated by the entropic term $\Pi_{\rm counter} = \kt cf$.
The contribution arising from the strand elasticity can be derived from the
elastic energy density that is proportional to the density of tensile-blobs :
\begin{equation}
F_{elastic}/k_{\rm B}T = (c/N)(N/m)(l/\xi) = \tau/(L_{\rm m}^2b)
\label{enelast}
\end{equation}
it varies as the $2/3$ power of concentration. As for the $\theta$-gel, we
assume that disentangled meshes are at equilibrium, $c = N/L_{\rm m}^3$. The
contribution of the elastic energy to the osmotic pressure
$cdF_{\rm elastic}/dc-F_{\rm elastic}$ is negative, and only differs from
$F_{\rm elastic}$ by a factor $-1/3$.  At the scaling level we thus write
\begin{equation}
\Pi_{\rm elastic}/k_{\rm B}T = -(c/N)(N/m)(L_{\rm m}/\xi) = -\tau/(L_{\rm m}^2 b)
\label{presselast}
\end{equation}
This is compensated by the counterion pressure $\Pi_{\rm count} = \kt cf$ at
the gel equilibrium
concentration :
\begin{equation}
c = {\tau^3\over N^2f^3b^3}
\label{eqconc}
\end{equation}
where the meshsize is
\begin{equation}
L_{\rm m} = b f N/\tau
 \label{meshsize}
\end{equation}
 It is easily checked that the mesh is stretched with respect to a
corresponding free necklace comprising
$N$ monomers, and as for a $\theta$-gel, the structure of the gel in
a poor solvent is not that of a
$c^\star$-solution. On the other hand
our description of the stretched necklace requires the necklace to be
shorter than a string of
thermal blobs as the pearl-necklace description remains valid only if most
monomers belong to pearls: $L<N\tau b$. Eq.(\ref{meshsize}) thus holds for
$\tau>f^{1/2}$; closer to the $\theta$-point,
the solvent quality becomes irrelevant and the structure is that of a
$\theta$-gel.
It can actually be checked that
the concentration smoothly crosses over with that
of the $\theta$-gel when $\tau = f^{1/2}$.

\begin{figure}
\begin{center}
\begin{minipage}{12cm}
\centerline{\epsfig{file=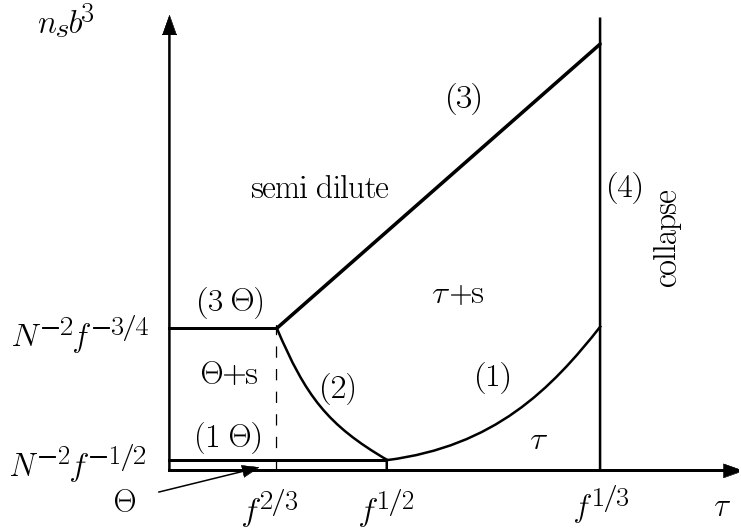,height=7cm}}
\vspace{10pt}
\caption{\label{fig:phase1} The phase diagram of the gel as a function of salt
  content and solvent quality, see text}
\end{minipage}
\end{center}
\end{figure}

For very poor solvents we know from the single necklace study that the
counterions are no longer free and condense on the pearls. The gel then
macroscopically collapses, this happens for
\begin{equation}
{l_{\rm B}\tau^3\over b f}>\left(\log {N^2 f^2 b^3\over \tau^3 }\right)^3
\label{gelcollapse}
\end{equation}
where the logarithm accounts for the counterion entropy at the onset of
condensation. Qualitatively, the gel is collapsed for $\tau>f^{1/3}$. In
summary, the poor solvent conditions are irrelevant close to the
$\theta$-point as long as $\tau <f^{1/2}$, the gel shrinks continuously for
intermediate solvents $f^{1/2}<\tau<f^{1/3}$ and undergoes collapse at $\tau
\sim f^{1/3}$ due to counterion condensation.  
This behavior is represented in FIG (\ref{fig:phase1}).

The compression modulus and the shear modulus of the equilibrium gel,
are proportional to the
counterion free energy density
and to the
elastic energy density respectively, and are of the same order at equilibrium:
\begin{equation}
G\propto K\propto \Pi_{\rm count} = \frac{\kt}{N^2 f^2} \left( \frac{
\tau}{b} \right)^3
\label{eqmod}
\end{equation}

When monovalent salt is added to the bulk solution, a Donnan equilibrium fixes
the inner salt concentration and the small-ion excess pressure. For salt
concentrations $n_{\rm s}$ smaller than the bound charge concentration $cf$,
salt is irrelevant, and the small ion excess pressure remains equal to $\kt
cf$. For $n_{\rm s}>cf$ in contrast, the small ions contribute a smaller inner
pressure excess $\Delta\Pi_{\rm ion} = \kt c^2f^2/(4 n_{\rm s})$.

This osmotic pressure difference is compensated by the elastic
contribution given by eq.(\ref{presselast}) at the equilibrium meshsize and
concentration:
\begin{equation}
L_{\rm m} = {N^{1/2} f^{1/2}\over \tau^{1/4} (n_{\rm s}b^3)^{1/4}}b \qquad
c = {\tau^{3/4} (n_{\rm s}b^3)^{3/4}\over N^{1/2}f^{3/2}b^3}
\label{csalt}
\end{equation}
At low salt concentration $n_{\rm s}b^3<N^{-2}f^{-1/2}(\tau/f^{1/2})^3$
these results crossover
the salt-free
results eqs.(\ref{eqconc},\ref{meshsize}), line (1) in the diagram 
FIG (\ref{fig:phase1}).
Close to the $\theta$-point,
$n_{\rm s}b^3<N^{-2}f^{-1/2}(\tau/f^{1/2})^{-5}$ the negative excluded
volume
becomes irrelevant and the results crossover to those obtained for a
$\theta$-gel,
line (2) in the diagram.
The weak screening  regime described by eq.(\ref{csalt}) is also bound at
high salt concentrations : the screening length due to the added salt
becomes shorter than the meshsize for $n_{\rm s}b^3 = \tau b^2 l_{\rm
B}^{-2}N^{-2}f^{-2}$; at
 higher
salt concentration the salt imposes the screening length, line (3) in the
diagram. On this line, the electrostatic interaction
between polymer strands in the mesh contributes as much as the small-ions
to the excess pressure. The
crosslinks are thus only marginally relevant. One easily checks that the
mesh size is indeed equal to the
size of a free strand of mass $N$. This means that, at the scaling level, there
is no difference between
the structure of the
gel and that of a semi-dilute solution above line (3). Lines (2) and (3)
meet at point
$B (\tau = f^{2/3}(l_{\rm B}/b)^{1/3}, n_{\rm s}b^3 =
f^{-4/3}N^{-2}(b/l_{\rm B})^{5/3})$. For poorer
solvents $\tau>f^{1/3}$,  the small ions condense on the pearls and the gel
collapses, line (4) in the
diagram; this line has not been drawn precisely, in particular the
condensation threshold weakly
depends upon the salt concentration $n_{\rm s}$.

\subsubsection{The affine deformation model}

The condition that the meshes only marginally overlap at swelling equilibrium
is now released and replaced by the more general condition of affine
deformation $c L_m^3/N =\alpha$ where the constant $\alpha$ is determined in
the preparation state. As explained earlier, the physical content of this
assumption is quite opposite to that of the $c^\star$-model. The
$c^\star$-model is nonetheless recovered in the special case $\alpha = 1$
corresponding to a semi-dilute solution where all temporary contact points
between different chains have been made permanent by the crosslinking
reaction.  In practice $\alpha>1$ reaches its highest value $N^{1/2}$ for
crosslinking in the neutral melt.  The elastic energy density now depends on
the preparation state through the parameter $\alpha$.
\begin{equation}
F_{el} = (c/N)(L_{\rm m}/\xi_{th}) = {\tau\alpha\over bL^2} = {\tau\over
b}\left(c\over N\right)^{2/3}\alpha^{1/3}
\label{elasticaffine}
\end{equation}
The equilibrium meshsize and concentrations have to be modified accordingly
and typically differ from those in the
$c^\star$-model by a power of $\alpha$.

\begin{figure}
\begin{center}
\begin{minipage}{12cm}
\centerline{\epsfig{file=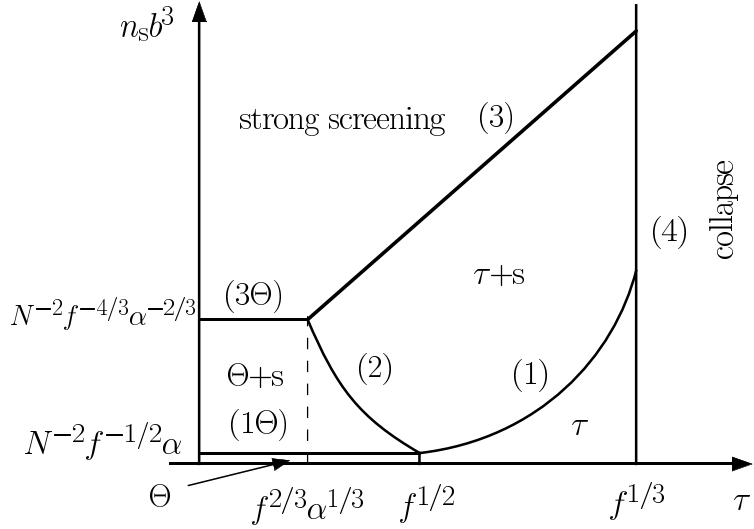,height=7cm}}
\vspace{10pt}
\caption{\label{fig:phase2} Phase diagram of the affine model. The general
  topology is preserved, see FIG.(\ref{fig:phase1}), but the axis are shifted
  and the ranges of the regimes are changed (see text).}
\end{minipage}
\end{center}
\end{figure}

\begin{figure}
\begin{center}
\begin{minipage}{12cm}
\centerline{\epsfig{file=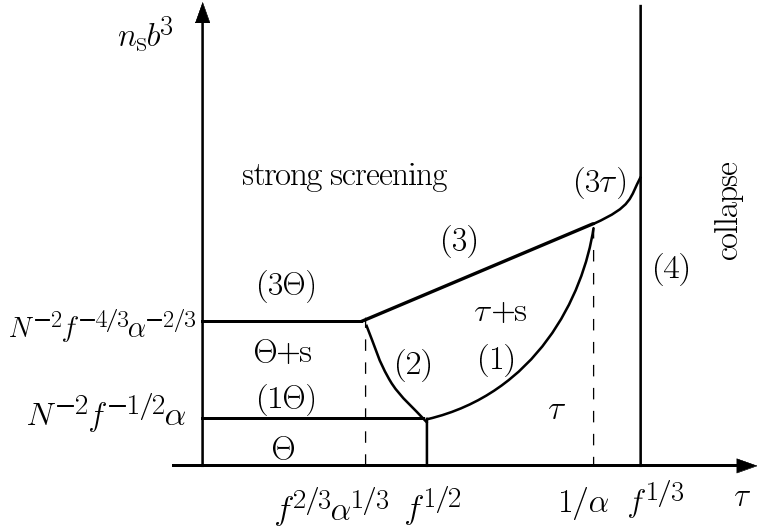,height=7cm}}
\vspace{10pt}
\caption{\label{fig:phase3} The phase diagram for poorer solvent, $\tau >
  1/\alpha$. The strong screening regime occures before the Donnan effect is
  relevant (see text).}
\end{minipage}
\end{center}
\end{figure}

For weakly entangled preparation states $1<\alpha<f^{-1/3}$ the topology of
the diagram of states presented in FIG.(\ref{fig:phase1}) is preserved. The
salt concentration on dividing lines (1), (1$\theta$) and (2) is increased by
a factor $\alpha$ whilst it is decreased by a factor $1/\alpha$ on line (3)
and by a factor $\alpha^{-2/3}$ on line (3$\theta$). The points where these
lines cross are shifted as shown in FIG.(\ref{fig:phase2}).  
As a consequence the
parameter range of the intermediate regimes $\theta + S$ and $\tau + S$ is
reduced. For somewhat more entangled preparation states
$f^{-1/3}<\alpha<f^{-1/2}$, lines (1) and (3) cross: For poor enough solvent
$\tau>1/\alpha$, strong screening occurs before the Donnan effect is relevant
as shown on FIG(\ref{fig:phase3}). 
In this poor solvent regime a new dividing line (3$\tau$) directly separates
the strongly screened region from the $\tau$ region where salt is irrelevant.
On this line $n_{\rm s} b^3 = b\tau^2/(\lb N^2f^2)$. There is direct contact
between the strong screening and the salt irrelevant region over the whole
$\tau$-range for $\alpha = f^{-1/2}$ (see figure 3). For higher values of
$\alpha$ the remaining lines (3) and (3$\theta$) no longer shift with
$\alpha$.

The expression of the gel equilibrium concentration and elastic modulus in
the various regimes
are summarized in table\ref{table}.
Let us stress that for $\alpha>1$, the gel structure on lines (3) is not
that of the solution
at the same concentration.
\begin{center}
\begin{minipage}{14cm}
\begin{table}
\begin{tabular}{|l|l|l|l|}
\hline
{\bf regime}   &{$\bf  L_{\rm m}/b$}     &${\bf  c b^3}$
& ${\bf G b^3/\kt}$                            \\
\hline
{$\bf \theta$}   & $N\sqrt{f}$    & $\alpha N^{-2}f^{-3/2}$ & $\alpha
N^{-2}f^{-1/2}$\\
\hline
{$\bf \theta+S$} &$\alpha^{1/5}f^{2/5}N^{3/5}(n_{\rm s}b^3)^{-1/5} $ &
$\alpha^{2/5}f^{-6/5}N^{-4/5}(n_{\rm s}b^3)^{3/5}
$ & $\alpha^{4/5}f^{-2/5}N^{-8/5}(n_{\rm s}b^3)^{1/5}$\\
\hline
${\bf \tau}$ & $Nf\tau^{-1}$ & $\alpha\tau^3 N^{-2} f^{-3}$ &
$\kt\alpha\tau^3 N^{-2} f^{-2} $\\
\hline
${\bf \tau + S}$ & $\alpha^{1/4}f^{1/2}N^{1/2}\tau^{-1/4}(n_{\rm s}b^3)^{-1/4}$ &
$\alpha^{1/4}f^{-3/2}N^{-1/2}\tau^{3/4}
(n_{\rm s}b^3)^{3/4}$ & $\alpha^{1/2}f^{-1}N^{-1}\tau^{3/2}(n_{\rm s}b^3)^{1/2}$\\
\hline
\end{tabular}
\vspace*{2mm}
\caption{Meshsize $L_{\rm m}$, gel concentration $c$ and modulus $G$ in the
various regimes}
\label{table}
\end{table}
\end{minipage}
\end{center}

\subsection{Shear modulus of non equilibrium gels}

When a small amount of water is added to the gel, the added amount imposes the
volume and the gel is not at swelling equilibrium.  In this case the
preparation state is of importance.  Most of the (elastic) properties depend
on the preparation state \cite{panyukov:98,edwards:88,deam:76}.  Rather
sophisticated theories have been developed for gels that have not reached
swelling equilibrium, but here we argue on a much simpler level.  The
counterion contribution to the osmotic pressure is given as above by the ideal
gas law. We can therefore conclude that the compression modulus is of the
order of $\kt c f$ with $c$ the actual gel concentration.  The shear modulus
on the other hand, depends on both the preparation state (crosslink
configuration) and the swelling state. For a sake of simplicity it is assumed
that the gel deforms affinely, and the preparation state is assumed to be an
undeformed isotropic state. Thus the measure of the affine deformation with
respect to the preparation size is determined by the ratio $\lambda =(c_{\rm
  p}/c)^{1/3}$.  where $c_{\rm p}$ and $c$ are the initial and final gel
concentrations, respectively. To estimate the elastic energy density stored in
the gel we take the semidilute solution at the same concentration $c$ where
the chains are stress free as a reference state.  At high enough
concentration, the elastic chains of the gel have gaussian statistics at large
length scale and behave as stretched Gaussian chains of blobs of radius
$\lambda R_{p}$. The modulus can be written as
\begin{equation}
G = \frac{\kt c}{N} \left( \frac{\lambda R_{\rm p}}{R}\right)^{2}
\label{noneqmod}
\end{equation}
where $R$ is the equilibrium radius of a chain in a semidilute solution at the
same concentration.  For moderate concentrations, $c < (\lb /b )^{1/2}f
\tau^{-1/2}b^{-3}$, several pearls are found in one correlation length
$\xi_{\rm c} = (\lb f^2/b)^{-1/4} \tau^{1/4}(cb^3)^{-1/2}$ and the chain
radius is given by equation \ref{1/2dilneck}.  We find then for the shear
modulus
\begin{equation}
G \propto \kt \left( \frac{\tau}{f^2}\right)^{1/4} c^{5/6}
\label{noneqgel}
\end{equation}
where the dependence upon the preparation state has been dropped. This
analysis is valid in the weak stretching limit where the gel strand is not
stretched. For more swollen gels, $c<c_1$, the radius of an elastic chain
$R_{c} = R_{\rm p}(c/c_{\rm p})^{-1/3}$, becomes larger than the radius of an
isolated necklace given in eq.(\ref{eqneck}). The crossover concentration
$c_1$ depends explicitly on the preparation state.  For $c<c_1$, the shear
modulus is dominated by the overstretched elastic chains and one should use
the elastic free energy of equation \ref{elasticaffine}
\begin{equation}
G =\kt c^{2/3} c_{\rm p}^{1/3}\tau R_{\rm p}/(Nb)
\label{Gstrong}
\end{equation}
Formally the modulus exhibits a sharp jump at the concentration $c_1$ within
the scaling picture.  Around $c_1$ a more careful consideration of the single
chain elasticity is needed. This could be done using a non gaussian elasticity
model such as the Langevin model for the chain of blobs, we are reluctant to
include it in our rough scaling analysis.

Within the affine deformation model the equilibrium swelling concentration
given by $G = cf$ depends explicitly on the preparation state, i.e.,
\begin{equation}
c_{eq} = c_{\rm p}R_{\rm p}^3\left({\tau\over Nf}\right)^3
\label{affineq}
\end{equation}
in contrast to the $c^\star$-model. The two approaches only agree if the
initial semi-dilute solution is cross-linked at all the contact points, where
the $c^\star$-theorem is indeed expected to apply. In that case, $c_{\rm
  p}R_{\rm p}^3 = N$ and eqs.(\ref{affineq},\ref{eqconc}) coincide. With this
preparation state, the shear modulus of a gel that has not reached equilibrium
reads:
\begin{equation}
G = k_{\rm B}T \left({c\over N}\right)^{2/3}{\tau\over b}
\label{cstarmodulus}
\end{equation}

Our results can  be tested experimentally by checking the final charge
fraction (or
temperature) dependence of the shear
modulus for identical  preparation states. Such experiments, where the
shear modulus was indeed
found to decrease with the charge fraction at fixed final density,
were performed in the group of J.Candau in Strasbourg
\cite{skouri:95}.One should note however that the experiments
were done with annealed
polyelectrolytes by monitoring the final charge with the pH of the
solution. Annealed polyelectrolytes, as shown recently
\cite{castelnovo}, if the solvent quality is too low can undergo a
sharp transition from a collapsed state to a stretched state
without any intermediate regime of stable necklace. Our results would
thus based on the pearl-necklace conformation would only be useful if
the solvent is not too poor which might be the case for polyacrylic
acid gels.

\section{Conclusion}

In this paper we have studied the elastic behavior of polyelectrolyte gels in
poor solvent which is markedly different from that of polyelectrolyte gels in
good solvent. The study of the single chain behavior already illustrates the
different physical behaviors. The stretching of polyelectrolyte chains in a
good solvent can be carried out via a simple blob analysis, as the undeformed
state of the chains can be described as an extended chain of electrostatic
blobs.  In a poor solvent, the undeformed chains are composed of pearls and
strings. The pearls act as a ''reservoir'' of monomers which can be pulled out
at relatively small forces. In the continuum limit the force increases gently
with the deformation. In the case of a low number of pearls, we expect length
jumps that correspond to single pearl unwinding at an imposed force. These
jumps are rounded by thermal fluctuations in many instances.

For a polyelectrolyte network, we do not expect any force jumps as the elastic
chains carrying the stress are much larger than a single mesh; the continuous
description is thus appropriate.  A gel at swelling equilibrium in a large
excess of $\theta$-solvent is known to swell\cite{barrat:92} under the effect
of the trapped counterion pressure. Most polyelectrolytes however have
hydrophobic backbones and the interplay between short range attractions and
electrostatics, that leads to the pearl-necklace conformation, is expected to
be important as well. Close to the $\theta$-point, for $\tau<f^{1/2}$, the
solvent quality is irrelevant and $c\propto f^{-3/2}$. Note that the crossover
between the $\theta$-solvent and the poor solvent behavior is different for
the equilibrium gel and for isolated chains ($\tau>f^{2/3}$). For intermediate
solvent quality $f^{1/2}<\tau<f^{1/3}$ the gel deswells and $c\propto
(\tau/f)^3$. When the solvent is very poor $\tau\sim f^{1/3}$, the counterions
condense on the pearls and the gel collapses.

When salt is added to the bulk solution, the gel shrinks. The various swelling
regimes are represented on the diagram of Fig.(\ref{fig:phase1}). If the salt
concentration is smaller than counterion concentration in the gel, the effect
of salt is negligible.  At higher salt concentration, the concentration in the
gel increases as $c\propto n_{\rm s}^{3/4}$.  At salt concentrations higher
than $\sim \tau/(N^2f^2)$, the gel has shrunk to the extent that junction
points no longer play any role, the properties of the gel are similar to those
of a semi-dilute polyelectrolyte solution (the details of this description
remain controversial).  In all the equilibrium swelling regimes the
compression and shear moduli are roughly proportional to each other and to the
small-ion excess pressure. The modulus is independent of the strength of the
electrostatic interactions, say of the Bjerrum length.

In the more realistic affine deformation model, the equilibrium swelling
depends upon the preparation state through the constant $\alpha = c L_{\rm
  m}^3/N$ measuring the degree of mesh entanglement. The affine deformation
constraint imposes that $\alpha$ remains constant during the swelling i.e.,
that the polymer content of a mesh volume remains constant and that meshes do
not disentangle.  The diagram of states is qualitatively similar to that
obtained form the $c^{*}$-model, meshsizes, the scaling laws for the
equilibrium concentration and the elastic modulus are shifted by a power of
$\alpha$ with respect to that obtained for the $c^\star$-gel that corresponds
to $\alpha = 1$. The results are summarized in table\ref{table}.  For higher
degrees of entanglement, $f^{-1/3}<\alpha<f^{-1/2}$, the diagram qualitatively
changes: For poor enough solvent $\tau>1/\alpha$ there is a direct separation
line between the salt irrelevant regime and the strong screening regime. For
$\alpha>f^{-1/3}$ this happens for any solvent quality.

A small amount of added solvent controls the gel volume, and the swelling
remains below the equilibrium swelling.  The compression and shear moduli are
no longer proportional to each other, even can show opposite variations with
the charge faction. Similar results where obtained previously for the
$\theta$-gels\cite{rubinstein:96} and observed in experiments\cite{skouri:95}.
We find a shear modulus $G\propto (\tau/f^{2})^{1/4}c^{5/6}$, that is shear
softening upon charge increase when all other parameters are kept constant.

Our picture relies on a scaling description that allows us to account for the
essential physics.  Some additional subtle effects may arise from the long
range character of the electrostatic interaction and its variation with the
salt concentration. For gels in a good solvent \cite{rubinstein:96} it has
been suggested recently \cite{wilder:98.1,wilder:98.2} that the elastic
modulus could depend on the strength of the interaction and the range of the
Debye - H\"uckel potential at least in some parameter range.  Our rough
approach would not capture those subtle effects.

\subsection*{Acknowledgments}
TAV acknowledges the financial support of the LEA. The warm hospitality of the
Institut Charles Sadron is gratefully appreciated.

\end{document}